# Results of the EURADOS international comparison exercise on neutron spectra unfolding in Bonner spheres spectrometry

J.M. Gómez-Ros[1,*], R. Bedogni[2], C. Domingo[3], J.S. Eakins[4], N. Roberts[5], R.J. Tanner[4]

[1]CIEMAT, Av. Complutense, 28040, Madrid, Spain

[2]INFN – LNF, Via E. Fermi n. 40, 00044 Frascati (Rome), Italy

[3]UAB, Physics Department, GRRI, 08193 Bellaterra, Spain

[4]United Kingdom Health Security Agency (UKHSA), Chilton, Didcot, Oxon OX11 0RQ, United Kingdom

[5]NPL, Hampton Road, Teddington, Middlesex TW11 0LW, United Kingdom

**\*Corresponding author:** José M. Gómez-Ros

**E-mail:** jm.gomezros@ciemat.es

**Telephone:** +34 913466237

**Fax:** +34 913466604

**Short running title:** BSS unfolding comparison exercise

**Abstract**

This paper summarizes the results obtained from an international comparison exercise on neutron spectra unfolding in Bonner spheres spectrometry, organized within the activities of EURADOS working group 6: computational dosimetry. Four realistic situations were considered: a medical accelerator, a workplace field, an irradiation room and a skyshine scenario. The reference solutions are presented, given in terms of idealized fluence-energy distributions and dose rates, along with details of their derivation using verified Monte Carlo codes. The wide variety of unfolded results that were submitted by the participants are then provided, with some shown to agree well with the reference solutions but others showing significant energy-dependent discrepancies. Finally, explanations for some of these discrepancies are proposed, along with suggested methods for how they might be improved.

## 1. Introduction

An "international comparison on neutron spectra unfolding in Bonner spheres spectrometry (BSS)" was organized within the activities of EURADOS working group 6 (WG6: *computational dosimetry*). The exercise considered an idealized $^3$He thermal neutron detector at the centre of a set of twelve polyethylene spheres of different radii, and the simulated measurements with them in four realistic exposure scenarios: a medical LINAC, a workplace field, an irradiation room, and skyshine irradiation (Gómez-Ros et al., 2018). There were two main objectives of the exercise: to provide validated training material, especially for novices in the field; and to warn users about frequent mistakes and potential pitfalls in the use of BSS unfolding.

The participants were provided with a similar amount of information about the irradiation scenarios as would be available in the real cases, and were asked to unfold the BSS data and provide their best estimates for: i) the neutron spectra with the appropriate normalization; ii) the integrated neutron fluences, $\Phi$, or fluence rates, $\dot{\Phi}$, in the thermal ($E < 0.4$ eV), epithermal (0.4 eV $< E <$ 0.1 MeV) and fast ($E >$ 0.1 MeV) energy regions; iii) the ambient dose equivalent, $H^*(10)$, or ambient dose equivalent rates, $\dot{H}^*(10)$; and iv) the fluence-averaged energy, $\bar{E}$, and the ambient dose equivalent-averaged energy, $\tilde{E}$, (ISO, 2001).

Twenty-one participants from fifteen countries submitted their solutions, using different unfolding codes, either standard or 'home-made'. In this paper, these solutions are compared with the reference spectra and integral quantities, and the different approaches adopted by the participants, as well as their most common mistakes, are categorized, presented and discussed.

## 2. Materials and Methods

For this comparison exercise, a BSS (Thomas and Alevra, 2002; Thomas, 2010) was considered that consisted of an idealized $^3$He thermal neutron detector and a set of polyethylene spheres of diameters 2″, 2″ + 1 mm Cd, 3″, 3.5″, 4″, 4.5″, 5″, 6″, 7″, 8″, 10″ and 12″. The $^3$He was assumed pure, and was a spherical volume enclosed in a 0.5 mm thick steel case with external diameter 33 mm, and hence internal diameter 32 mm (Gómez-Ros et al., 2018). The decision to use an ideal, or perfect, Bonner sphere set, rather a model of a real set, was intended to produce a pure unfolding intercomparison. Uncertainties in the precise geometry connected to the realistic detector and cable were removed from the exercise. It also meant that the responses of the Bonner spheres were truly isotropic.

The response matrix (Figure 1) of the BSS was obtained using the Monte Carlo code MCNPX 2.7 (Pelowitz, 2011), with the ENDF/B-VII neutron cross-section library and thermal neutron S($\alpha$,$\beta$) cross-section data for polyethylene (Chadwick et al., 2006) employed to calculate the number of $^3$He(n,p) reactions in the $^3$He-filled volume. The same procedure was used to calculate the BSS counts for the bare detector and for the twelve polyethylene spheres in the different irradiation scenarios. The reference neutron spectra, i.e. the neutron spectra at the measurement points, were calculated using track length fluence tallies defined on small volumes around those points in the absence of the spheres, and therefore corresponded to the unperturbed neutron fields. In all the cases, a sufficiently large number of histories have been generated to obtain simulation uncertainties lower than 0.2% for the integral quantities $\Phi$, $H^*(10)$, $\bar{E}$ and $\tilde{E}$, described in the previous section.

As was described in a previous paper (Gómez-Ros et al., 2018), four representative exposure scenarios were proposed (Figure 2). They were the following:

*2.1. Medical LINAC accelerator*

A 25 MV medical LINAC accelerator (GE Saturne 43), operated in photon mode. The accelerator is situated at the centre of a 7×7×3 m$^3$ treatment room with concrete walls, as indicated in Figure 2a. The accelerator head is mounted vertically, with the top of the target located 2 m above the floor. A 40×40×40 cm$^3$ water phantom with 1.5 cm thick PMMA walls is located in the beam, with its upper surface at 90 cm from the top of the target. The distance from the target to the isocentre is 1 m. Two measurement positions have been considered: one at the entrance of the maze (P1 in Figure 2a), and the other at 1 m from the isocentre (P2 in Figure 2a). The BSS counts for both points obtained from pure simulation were only affected by fluctuations due to simulation statistics, which do not consider the measurement uncertainties that would be encountered in a real case. To mimic realistic counts from an instrument, the simulated ones were randomly perturbed using a Poissonian distribution to simulate a counting uncertainty of 2%, a typical value for this type of measurements.

The participants were provided with the BSS response matrix and the BSS counts normalized to an absorbed dose to water of 1 Gy at the isocentre. Because the plane of the bunker and the LINAC type were the only information about the facility that was available to the participants, the main difficulty for them was assembling the 'pre-information' for the unfolding code. The following two possible approaches could be used:

- For unfolding codes requiring a default spectrum, a simplified Monte Carlo simulation with a 25 MeV electron pencil beam impinging on a high-Z target (for instance W, Ta, Au) can be performed. The room geometry can be included easily in the Monte Carlo model to calculate default spectra at the appropriate positions.

- For unfolding codes relying on parameterized spectra, the initial guess can be built by linearly combining a Maxwellian thermal component, a simplified $1/E$ epithermal continuous distribution, and an evaporation spectrum. The fractions of thermal, epithermal and evaporative neutrons, and the evaporation temperature, could be chosen to produce spectra with dominant thermal or fast contributions, according to the location of the measurement point (IAEA, 1990; IAEA, 2001).

*2.2. $^{241}$Am-Be based workplace field*

A simulated workplace field consisting of an ISO $^{241}$Am-Be source (ISO, 2001) suspended in a stainless-steel tube, which is clad with a lead shield, and partially moderated with a water-filled container. The steel tube extends from the floor to the ceiling of a 2.5×5.0×7.75 m$^3$ room with wooden panels covering all surfaces and neutron absorbing material placed behind the panels. The measurement point is located at 1 m distance from the source, at the same height (1.25 m). A moderating water-filled container of approximately 50 cm depth is situated adjacent to the steel tube, and in-line with the measurement point. The container is sufficiently wide to cover the solid-angle subtended at the source by the largest Bonner sphere, and is supported by a 2 cm thick wooden table at a height that gives equal coverage in the vertical direction. The statistical uncertainty of the neutron counts is 4%.

As in the previous scenario (i), a simplified Monte Carlo calculation of the exposure conditions can easily be performed to obtain a guess spectrum. Alternatively, a suitable parameterized spectrum may also be produced by linearly combining a Maxwellian thermal component, a simplified $1/E$ epithermal continuous distribution, and a fast spectrum that is similar to that of the $^{241}$Am-Be source. An evaporation spectrum with a temperature corresponding to about 2 MeV could also even be adopted.

*2.3. Irradiation room with a moderated $^{241}$Am-Be source*

An irradiation room scenario, featuring an ISO $^{241}$Am-Be source in the centre of a 10 cm radius iron sphere, located in the centre of a 8x8x8 m$^3$ irradiation room with 50 cm thick concrete walls, floor and ceiling. The measurement point is positioned at the same height as the source, at 4 m distance along the diagonal of the room's horizontal plane. The uncertainty of the simulated measurements is 2%.

Because the information available to the participants included the source type and the dimensions and materials of all components of the room, obtaining an *a priori* spectra or a parametric approach was easily feasible.

*2.4. Skyshine field around a nuclear plant*

The skyshine field at 100 m from a nuclear plant, which was modelled as a cylindrical building of 10 m height and 10 m radius with concrete walls. The roof is made from thin concrete tiles. The source term is represented by a point-like ($\alpha$,n) neutron source located in the middle of the building. The total activity of the source is unknown to the participants. As the walls substantially attenuate the direct field, the overall field at the measurement point is dominated by air-scattered neutrons. The Bonner spheres were placed with their centres 1.5 m above the ground. The uncertainty of the simulated measurements is 5%.

Assembling a Monte Carlo based spectrum *a priori* would be quite difficult for this scenario, described in section 3.4, because advanced biasing techniques were required to describe the air-scattered field at 100 m distance with sufficient precision. The parametric approach is therefore easier in this case.

**3. Results and discussion**

In total, sixty-four solutions from twenty participants were received. In order to preserve the participants' anonymity, they are randomly identified by letters (a) to (t) in the following analyses; the same letter is maintained for each of the solutions provided for the different scenarios by a given participant. A summary of the unfolding codes and scenarios that were attempted is presented in Table 1, along with the 'pre-information method' used by each participant to derive their guess spectra. As can be seen, the UMG (MAXED, GRAVEL) package (Reginatto et al., 2002; Matzke, 2003) was the most widely used code (11 participants), followed by FRUIT (Bedogni et al., 2007) (3 participants). The following codes were each used by a single participant: B-UNCLE (Pola et al., 2020), GRUPINT / ANGELO / ZOTT99 (Muir, 1999; Kodeli and Sartori, 2010; Trkov et al., 2017), MSITER / MIEKE (Szondi, 1999; Zsolnay et al., 1999) and WinBUGS (Mazrou and Bezoubiri, 2018).

In addition to the solutions derived using the above software packages, three participants utilized different approaches for their unfolding. Participant (*s*) used a hybrid Tykhonov method, where a regularization term was added to a least-squares criterion (Besida, 2005). Participant (*t*) used analysis software written in C++, utilizing the ROOT software libraries (Brun and Rademakers, 1997) to analyse the singular value decomposition of the response matrix (Rust, 1998). In that method, the data vector was created by using a cubic spline through the provided BSS data, and sampling the spline at equal intervals to generate sufficient components to equal the number of energy bins requested in the problem.

A detailed description of the unfolding algorithms used by each software can be found in the cited references. In addition, good surveys on the unfolding techniques used in neutron spectrometry have been already published (Reginatto, 2010; Barros et al., 2014). Essentially, these methods are based either on iteratively variation of an initial spectrum according to the numerical values of the sphere counts, either on a non-linear fitting of a parameterised spectrum, described by a limited number of physically meaningful parameters. In all the cases, some prior information or pre-information, based on physical considerations, needs to be provided because the unfolding process is an under-determined with infinite mathematical solutions for a given response matrix and a set of measurement counts (Reginatto, 2010). This pre-information will be an initial guess spectrum, usually obtained by Monte Carlo simulation (as it is indicated in Table 1, with MCNP5, MCNPX, MCNP6 codes) or from published references (IAEA, 1990; IAEA, 2001), or a set of initial values for the parametrised spectrum parameters, according to the expected physical characteristics.

The sixty-four solutions from the participants are commented, scenario by scenario, in Sections 3.1 to 3.4. The participants' spectra, normalized to the total fluence (unit spectra) and plotted in terms of lethargy ($E\Phi_E$), are shown in Figures 3-7 compared against the reference spectra and then discussed based on the qualitative criteria given in section 3.5. In addition, the relative difference (%) between the integral quantities' values provided by the participants and those corresponding to the reference simulated spectra, listed in Table 2, have been calculated to quantify the discrepancy of the unfolded spectra, i.e., for any of the integral quantities, *Q*:

$$\Delta(\%) = \frac{Q_{(participant)} - Q_{(reference)}}{Q_{(reference)}} \times 100 \qquad (1)$$

The values of the relative differences for the different scenarios are listed in tables 3-7. A graphical representation these values compared with the reference ones can be found in Supplementary Figures 1-5. The complete results submitted by participants are shown in Supplementary Figures 6 through 25.

*3.1. Medical LINAC*

The solutions proposed by the participants for point P1 (at the entrance of the maze) are shown in Figure 3. Five of those solutions were not correct, i.e., they significantly differ from the reference spectrum (expected to be measured): for solution (*g*) the thermal peak is broader and less intense than expected, and the maximum of the fast neutron peak is at 0.1 MeV; for solution (*h*) the thermal neutron peak is slightly overestimated; for solution (*n*) the fast neutron peak is slightly overestimated; for solution (*s*) the thermal peak is broader and less intense than expected, and the fast peak is broader than expected; and for solution (*t*) the thermal peak is broader and less intense than expected.

For the LINAC point P2 (1 m from the isocentre), all the participants' unfolded spectra are shown in Figure 4. As before: in solution (*g*) the thermal peak is broader and less intense than expected, and the fast neutron peak is also broader and displaced to lower energies; in solution (*h*) the fast neutron peak is slightly overestimated; in solution (*n*) the thermal peak is overestimated, and the fast neutron peak is slightly underestimated; in solution (*s*) the thermal peak is broader and less intense than expected, and the fast peak is slightly underestimated; and in solution (*t*) the thermal peak is broader and less intense than expected, and the fast peak slightly underestimated.

Tables 3 and 4 respectively show the relative differences between participants and reference values of spectrum-integrated quantities for point P1 (at the entrance of the maze) and point P2 (1 m from the isocentre). Those quantities are: fluence rate $\dot{\Phi}$, total as well as in broad energy intervals ($E<0.4$ eV, $0.4$ eV $< E < 0.1$ MeV, $E > 0.1$ MeV), ambient dose equivalent rate, $\dot{H}^*(10)$, fluence-averaged-energy, $\bar{E}$, and $H^*(10)$-averaged-energy, $\tilde{E}$.

*3.2. Workplace field*

All the participants' unfolded spectra are shown in Figure 5, where five incorrect solutions could be identified: for solution (*h*) the thermal peak is underestimated, and the fast peak is overestimated and displaced to higher energies; for solution (*j*) the thermal peak is slightly underestimated and the fast peak slightly overestimated; for solution (*l*) the thermal peak is slightly overestimated; for solution (*s*) the thermal peak is broader, less intense than expected, and displaced to higher energies; and for solution (*t*) the thermal peak is also broader, less intense than expected, and displaced to higher energies. Table 5 shows the relative differences between the integral quantities provided by the participant and the reference.

*3.3. Irradiation room with a radionuclide source*

All the participants' unfolded spectra are shown in Figure 6, and six of those solutions are shown in more detail in Figure 4c. In that figure: for solutions (*a*), (*b*) and (*c*) the fast peak is overestimated and its maximum is slightly displaced to higher energies; solution (*p*) is completely unrealistic; and for solutions (*s*) and t) the thermal peak is broader and less intense than expected, and displaced to higher energies. Table 6 shows the relative differences between the integral quantities provided by the participant and the reference.

*3.4. Skyshine scenario*

As discussed in Section 2, this scenario was especially difficult to assemble consistent *a priori* information for the unfolding code. Consequently, around 70% of the solutions were not correct, as it can be seen in Figure 7. Solutions, (*a*), (*b*), (*c*), (*h*), (*k*), (*l*), (*m*), (*r*), (*s*) and (*t*) are seen to overestimate the fast neutron peak. In addition: solutions (*k*) and (*l*) show an unrealistic double peak near the thermal region, and both solutions also underestimate the epithermal contribution; for solution (*r*) the thermal peak is broader and less intense than expected and displaced to higher energies, and the epithermal contribution is underestimated; and for solution (*s*) the thermal peak is broader and less intense than expected, and displaced to higher energies. Table 7 shows the relative differences between the integral quantities provided by the participant and the reference.

3.5    *Analysis of the solutions*

There could be many possible causes for anomalous or distorted solutions. Obvious explanations might include inadequate use of the unfolding code, or inacurate formulation of the information used as input for the unfolding code, for example in terms of either the guess spectrum or selected parameters. Because the neutron unfolding problem is underdetermined, in general it has an infinite number of possible mathematical solutions (Reginatto, 2010). Therefore, some previous information (i.e. pre-information) is required: for parametric unfolding codes, an initial estimation of the key parameters is needed; for nonparametric codes, a plausible guess spectrum must be supplied. In both cases, those guess values need to be sufficiently close to the actual values, otherwise the unfolding procedure can converge to a mathematically correct, but physically unrealistic solution.

In many cases, however, anomalous results could potentially have been self-identified by the participants had they performed simple verification tests or considered the likely physical plausibility of their results. Example such tests could be:

- Plotting the unfolded spectrum. In some cases, unrealistic results, such as solution (p) in Figure 6e, could then be identified by eye. Moreover, some participants provided spectra with negative fluence values in some energy bins. This is also something that can be easily recognized in the plots.

- Comparing the Reference BSS counts with those obtained by applying the unfolded spectrum to the response matrix. However, it is remarked that whilst the compatibility of these values is a necessary condition, it does not on its own guarantee the correctness of the result because the BSS unfolding problem is an underdetermined process. Therefore, the unfolded spectrum could still be incorrect if guess spectrum or default parameters are unrealistic, for instance a flat-in-lethargy $1/E$ function. This is the reason why the values of fluence integrated over a large energy domain may tend to be correct even when the spectrum shape has not been correctly determined. On the contrary, the values of quantities that depend heavily on the energy distribution, such as $H^*(10)$-averaged energy, fluence-averaged energy and $\dot{H}^*(10)$, tend to exhibit large deviations from reference values for those spectra.

- Comparing the unfolded spectrum with those reported in literature (e.g. IAEA, 1990; IAEA, 2001) for similar scenarios.

As expected, the discrepancy between the unfolded spectra and the reference one is more significant for those problems where less detailed pre-information was available, i.e. scenarios i) LINAC (40% of solutions were poor) and iv) skyshine (70% of solutions were poor). Nevertheless, around one third of the submitted solutions were also incorrect for scenarios ii) and iii), for which the problem was much better specified in terms of source type and the sizes and materials of all objects in the room.

Some of the anomalous solutions were submitted by participants who used "self-made" or "non-standard" codes. While "standard" unfolding codes are referenced and benchmarked in many situations, thus providing a reasonable guarantee of their ability to give correct solutions whenever correctly used, "self-made" and "non-standard" codes would require an initial verification process about their capability to correctly solve the problem and to converge to a correct solution.

**4. Conclusions**

Although computer codes for unfolding BSS data have been in use for a long time, their application remains a complex task that needs some technical skills and experience in the specific field. This is true regardless of whether or not the unfolding problem itself is inherently complex. Indeed, the results of this exercise suggest that difficulties exist in general during attempts to correctly perform BSS unfolding tasks. This difficulty, expressed in terms of the proportion of incorrect solutions, was observed to have a base value of about 30% even in the simple scenarios considered here (ii and iii), for which almost complete prior information was available. The difficulty then tends to increase as the pre-information decreases.

Preparing reliable pre-information for BSS unfolding is crucial. Because the neutron unfolding problem is underdetermined, even validated unfolding codes may lead to inaccurate solutions if the code is fed non-realistic information.

Unfolding codes, even the most sophisticated and documented ones, cannot be used without the required expertise of the user if accurate results are to be obtained. To that end, such users should: 1) have sufficient physical knowledge of the radiation environment to enable him/her to estimate the appropriate likely characteristics of that field; 2) have sufficient mathematical capabilities to translate that physical knowledge into *a priori* information that may be suited for the specific unfolding code being used; and 3) have sufficient experience to enable him/her to correctly judge the results of the unfolding process for likely plausibility and accuracy.

All the data corresponding to this EURADOS exercise on BSS unfolding: description, response matrix, BSS counts and reference spectra are available as supplementary files.

ACKNOWLEDGEMENTS

This work is been partially supported by EURADOS, within the activities of Working Group 6: Computational Dosimetry. The organizers of the exercise wish also to thanks all the participants for their valuable contributions: John Paul Archambault (NRC, Canada), Jovica Atanackovic (Ontario Power Generation, Canada), Thierry Buchillier (Lausanne University Hospital, Switzerland), Jorge Carelli (ARN, Argentina), Thomas Donaldson McLean (LANL, USA), Silva Everton (IRD, Brazil), Hideki Harano (NMIJ, Japan), Sang In Kim (KAERI, Republic of Korea), Jungho Kim (KRISS, Republic of Korea), Ivan-Alexander Kodeli (Jozef Stefan Inst., Slovenia), Bor Kos (Jozef Stefan Inst., Slovenia), Seung Kyu Lee (KAERI, Republic of Korea), Imma Martínez-Rovira (UAB, Spain), Akihiko Masuda (NMIJ, Japan), Tetsuro Matsumoto (NMIJ, Japan), Hakim Mazrou (CNRA, Algerie), Ignacio Menchaca (ARN, Argentina), Valeria Monti (INFN, Italy), Thiem Ngoc Le (VINATOM, Vietnam), Quynh Ngoc Nguyen (VINATOM, Vietnam), Tamás Pázmándi (Centre for Energy Research, Hungary), Vladimir Radulović (Jozef Stefan Inst., Slovenia), Dario Rastelli (Raylab, Italy), Maite Romero-Expósito (UAB, Spain), Sujoy Sen (Indira Gandhi Centre for Atomic Research, India), Olivier Van Hoey (SCK-CEN, Belgium).

**TABLES:**

**Table 1.** Summary of participants' unfolding codes, the scenarios they solved, and the pre-information method that was adopted.

| participant | unfolding method | LINAC | workplace | calibration room | skyshine | pre-information |
|---|---|---|---|---|---|---|
| a | B-UNCLE | x | x | x | x | not clearly indicated |
| b | FRUIT | x | x | x | x | choice of parametric model |
| c | FRUIT | x | x | x | x | choice of parametric model |
| d | FRUIT | x | x | x | x | missing information |
| e | GRUPINT, ANGELO, ZOTT99 | x | x | x | x | MCNP6 |
| f | UMG 3.3 | x |  | x |  | MCNP6 |
| g | UMG 3.3 | x |  |  |  | default spectrum from literature |
| h | UMG 3.3 | x | x | x | x | MCNPX 2.5 |
| i | UMG 3.3 |  | x | x | x | MCNP6 |
| j | UMG package: MXD_FC33 |  | x | x |  | MCNP6 |
| k | MAXED | x | x | x | x | problem dependent |
| l | GRAVEL | x | x | x | x | problem dependent |
| m | MXD_FC33 and IQU_FC33 | x | x | x | x | problem dependent |
| n | MAXED | x | x | x | x | MCNP5 |
| o | MAXED / UMG |  |  | x |  | MCNP5 |
| p | MAXED 2000 |  |  | x |  | not clearly indicated |
| q | MSITER / MIEKE |  | x | x |  | MCNP5 |
| r | WinBUGS | x | x | x | x | choice of parametric model |
| s | basic Tykhonov method | x | x | x | x | none |
| t | self-made | x | x | x | x | none |

**Table 2:** Integral quantities values for the reference spectra: fluence, Φ, (per unit of absorbed dose at the isocentre) and ambient dose equivalent, $H^*(10)$, for the LINAC; fluence rate, $\dot{\Phi}$, and ambient dose equivalent rate, $\dot{H}^*(10)$, for the other three scenarios; fluence-averaged energy, $\bar{E}$; $H^*(10)$-averaged energy, $\tilde{E}$. Relative uncertainties are always lower than 0.2%.

| quantity | scenario | | | | |
|---|---|---|---|---|---|
| | LINAC (P1) | LINAC (P2) | workplace | calibration room | workplace |
| Φ, $\dot{\Phi}_{total}$ | 6.79x10$^6$ cm$^{-2}$Gy$^{-1}$ | 2.27x10$^7$ cm$^{-2}$Gy$^{-1}$ | 22.9 cm$^{-2}$s$^{-1}$ | 6.92 cm$^{-2}$s$^{-1}$ | 643 cm$^{-2}$h$^{-1}$ |
| Φ, $\dot{\Phi}_{E<0.4eV}$ | 3.05x10$^6$ cm$^{-2}$Gy$^{-1}$ | 6.00x10$^6$ cm$^{-2}$Gy$^{-1}$ | 10.6 cm$^{-2}$s$^{-1}$ | 2.58 cm$^{-2}$s$^{-1}$ | 169 cm$^{-2}$h$^{-1}$ |
| Φ, $\dot{\Phi}_{0.4eV<E<0.1MeV}$ | 1.88x10$^6$ cm$^{-2}$Gy$^{-1}$ | 5.04x10$^6$ cm$^{-2}$Gy$^{-1}$ | 6.36 cm$^{-2}$s$^{-1}$ | 1.84 cm$^{-2}$s$^{-1}$ | 264 cm$^{-2}$h$^{-1}$ |
| Φ, $\dot{\Phi}_{E>0.1MeV}$ | 1.86x10$^6$ cm$^{-2}$Gy$^{-1}$ | 1.17x10$^7$ cm$^{-2}$Gy$^{-1}$ | 5.98 cm$^{-2}$s$^{-1}$ | 2.51 cm$^{-2}$s$^{-1}$ | 210 cm$^{-2}$h$^{-1}$ |
| $H^*(10)$, $\dot{H}^*(10)$ | 6.34x10$^8$ pSv Gy$^{-1}$ | 4.03x10$^9$ pSv Gy$^{-1}$ | 2.32x10$^3$ pSv s$^{-1}$ | 8.82x10$^2$ pSv s$^{-1}$ | 7.75x10$^4$ pSv h$^{-1}$ |
| $\bar{E}$ | 0.233 MeV | 0.533 MeV | 0.503 MeV | 0.510 MeV | 0.582 MeV |
| $\tilde{E}$ | 0.940 MeV | 1.170 MeV | 2.005 MeV | 1.579 MeV | 1.930 MeV |

**Table 3:** Relative difference (%) between the integral quantities values provided by the participants and the corresponding reference values for the LINAC scenario (point P1): fluence, $\Phi$, (total and by three energy groups), ambient dose equivalent rate, $\dot{H}^*(10)$, fluence-averaged-energy, $\bar{E}$, and $H^*(10)$-averaged-energy, $\tilde{E}$.

| solution | $\Phi_{total}$ | $\Phi_{E<0.4eV}$ | $\Phi_{0.4eV<E<0.1MeV}$ | $\Phi_{E>0.1MeV}$ | $H^*(10)$ | $\bar{E}$ | $\tilde{E}$ |
|---|---|---|---|---|---|---|---|
| a | 3% | 3% | 21% | -14% | -7% | -2% | 12% |
| b | 4% | 3% | 23% | -14% | -10% | 13% | 34% |
| c | 4% | 3% | 22% | -13% | -9% | 11% | 30% |
| d | 1% | -2% | 19% | -11% | -5% | 5% | 15% |
| e | 4% | <1% | 27% | -12% | -9% | 18% | 39% |
| f | 3% | 1% | 8% | <1% | -2% | 5% | 11% |
| g | 2% | -9% | 23% | -2% | -9% | 2% | 13% |
| h | 5% | 14% | -1% | -5% | -1% | 18% | 29% |
| i | - | - | - | - | - | - | - |
| j | - | - | - | - | - | - | - |
| k | 3% | 1% | 17% | -7% | -7% | 29% | 48% |
| l | -6% | -8% | 6% | -15% | -16% | 15% | 31% |
| m | 2% | -1% | 18% | -8% | -4% | -3% | 5% |
| n | 1% | -7% | 4% | 12% | 8% | -9% | -17% |
| o | - | - | - | - | - | - | - |
| p | - | - | - | - | - | - | - |
| q | - | - | - | - | - | - | - |
| r | 3% | 1% | 27% | -16% | -10% | 8% | 29% |
| s | 2% | -6% | 29% | -11% | -9% | 61% | 98% |
| t | -1% | -6% | 18% | -13% | -11% | 3% | 17% |

**Table 4:** Relative difference (%) between the integral quantities values provided by the participants and the corresponding reference values for the LINAC scenario (point P2): fluence, $\Phi$, (total and by three energy groups), ambient dose equivalent rate, $\dot{H}^*(10)$, fluence-averaged-energy, $\bar{E}$, and $H^*(10)$-averaged-energy, $\tilde{E}$.

| solution | $\Phi_{total}$ | $\Phi_{E<0,4eV}$ | $\Phi_{0.4eV<E<0.1MeV}$ | $\Phi_{E>0.1MeV}$ | $H^*(10)$ | $\bar{E}$ | $\tilde{E}$ |
|---|---|---|---|---|---|---|---|
| a | <1% | -5% | 15% | -4% | 2% | 1% | 2% |
| b | 1% | -4% | 17% | -3% | 1% | <1% | 2% |
| c | 1% | -4% | 13% | -2% | 2% | -1% | -1% |
| d | <1% | -4% | 3% | 1% | 4% | -4% | -7% |
| e | 3% | -2% | 25% | -3% | -4% | 27% | 38% |
| f | 1% | -4% | <1% | 5% | 3% | 10% | 8% |
| g | 2% | -11% | 20% | <1% | -6% | -6% | <1% |
| h | 1% | -5% | -10% | 9% | 14% | 44% | 31% |
| i | - | - | - | - | - | - | - |
| j | - | - | - | - | - | - | - |
| k | 2% | -4% | -4% | 8% | 3% | 25% | 25% |
| l | -5% | -11% | -10% | <1% | -5% | 12% | 12% |
| m | <1% | -7% | 15% | -2% | <1% | -3% | -2% |
| n | 1% | 42% | 30% | -33% | -36% | -54% | -31% |
| o | - | - | - | - | - | - | - |
| p | - | - | - | - | - | - | - |
| q | - | - | - | - | - | - | - |
| r | 1% | -4% | 17% | -4% | <1% | -5% | -3% |
| s | 1% | -15% | 23% | 1% | -1% | 30% | 35% |
| t | 1% | -16% | 18% | 3% | 1% | 15% | 15% |

**Table 5:** Relative difference (%) between the integral quantities values provided by the participants and the corresponding reference values for the workplace scenario: fluence rate, $\dot{\Phi}$, (total and by three energy groups), ambient dose equivalent rate, $\dot{H}^*(10)$, fluence-averaged-energy, $\bar{E}$ and $H^*(10)$-averaged-energy, $\tilde{E}$.

| solution | $\Phi_{total}$ | $\Phi_{E<0.4eV}$ | $\Phi_{0.4eV<E<0.1MeV}$ | $\Phi_{E>0.1MeV}$ | $H^*(10)$ | $\bar{E}$ | $\tilde{E}$ |
|---|---|---|---|---|---|---|---|
| a | -1% | 2% | -5% | -2% | -5% | -33% | -31% |
| b | -1% | -1% | 7% | -10% | -10% | -36% | -31% |
| c | -2% | -7% | 16% | -11% | -13% | -31% | -24% |
| d | -2% | -4% | 11% | -10% | -11% | -36% | -29% |
| e | -2% | -7% | 10% | -5% | -10% | 4% | 15% |
| f | - | - | - | - | - | - | - |
| g | - | - | - | - | - | - | - |
| h | -1% | -15% | -5% | 29% | 27% | 97% | 55% |
| i | -3% | -11% | 3% | 5% | -1% | 14% | 11% |
| j | -8% | -30% | -21% | 45% | 41% | 92% | 26% |
| k | -2% | 3% | <1% | -11% | -6% | -34% | -31% |
| l | 4% | 8% | 2% | -2% | -1% | -26% | -23% |
| m | -1% | -1% | 4% | -6% | -7% | <1% | 7% |
| n | -4% | -13% | 6% | 1% | -4% | -14% | -15% |
| o | - | - | - | - | - | - | - |
| p | - | - | - | - | - | - | - |
| q | -1% | <1% | 2% | -7% | -8% | -22% | -17% |
| r | -1% | -2% | 6% | -8% | -9% | -21% | -15% |
| s | -1% | -7% | 13% | -6% | -5% | 32% | 50% |
| t | -4% | -6% | -4% | -1% | -5% | -26% | -26% |

**Table 6:** Relative difference (%) between the integral quantities values provided by the participants and the corresponding reference values for the calibration room scenario: fluence rate, $\dot{\Phi}$, (total and by three energy groups), ambient dose equivalent rate, $\dot{H}^*(10)$, fluence-averaged-energy, $\bar{E}$ and $H^*(10)$-averaged-energy, $\tilde{E}$.

| solution | $\Phi_{total}$ | $\Phi_{E<0.4eV}$ | $\Phi_{0.4eV<E<0.1MeV}$ | $\Phi_{E>0.1MeV}$ | $H^*(10)$ | $\bar{E}$ | $\tilde{E}$ |
|---|---|---|---|---|---|---|---|
| a | -2% | -4% | <1% | -1% | 6% | -19% | -25% |
| b | -2% | -4% | 1% | -1% | 6% | -20% | -25% |
| c | -1% | -4% | 5% | -4% | 4% | -20% | -23% |
| d | <1% | -2% | <1% | 3% | 4% | 4% | 1% |
| e | -1% | -3% | -8% | 7% | 5% | -10% | -16% |
| f | <1% | -6% | 5% | 2% | 2% | 3% | <1% |
| g | - | - | - | - | - | - | - |
| h | 1% | 4% | -1% | 1% | 2% | 4% | 4% |
| i | -7% | -30% | 12% | 4% | 1% | 9% | <1% |
| j | 2% | 2% | 2% | 1% | 1% | -2% | -1% |
| k | -1% | -2% | -3% | 3% | 5% | -5% | -10% |
| l | <1% | -1% | -2% | 3% | 6% | 1% | -5% |
| m | 1% | 1% | -2% | 2% | 4% | 12% | 10% |
| n | -1% | -16% | 12% | 3% | 1% | 3% | <1% |
| o | -1% | -5% | 4% | -2% | -1% | <1% | <1% |
| p | * | * | * | * | * | * | * |
| q | <1% | -3% | 1% | 2% | 4% | 2% | -2% |
| r | -1% | -3% | <1% | <1% | 5% | -16% | -20% |
| s | -1% | -11% | 8% | 2% | 3% | 12% | 9% |
| t | -3% | -15% | 2% | 5% | 4% | -5% | -12% |

**Table 7:** Relative difference (%) between the integral quantities values provided by the participants and the corresponding reference values for the skyshine scenario: fluence rate, $\dot{\Phi}$, (total and by three energy groups), ambient dose equivalent rate, $\dot{H}^*(10)$, fluence-averaged-energy, $\bar{E}$ and $H^*(10)$-averaged-energy, $\tilde{E}$.

| solution | $\dot{\Phi}_{total}$ | $\dot{\Phi}_{E<0.4eV}$ | $\dot{\Phi}_{0.4eV<E<0.1MeV}$ | $\dot{\Phi}_{E>0.1MeV}$ | $\dot{H}^*(10)$ | $\bar{E}$ | $\tilde{E}$ |
|---|---|---|---|---|---|---|---|
| a | -3% | -15% | 3% | -1% | 5% | -4% | -11% |
| b | -2% | -14% | -1% | 6% | 12% | -29% | -37% |
| c | -2% | -14% | -1% | 6% | 12% | -28% | -37% |
| d | -3% | -9% | -8% | 7% | 7% | -2% | -12% |
| e | -2% | -13% | -2% | 9% | 6% | -9% | -16% |
| f | - | - | - | - | - | - | - |
| g | - | - | - | - | - | - | - |
| h | 0% | -25% | -15% | 39% | 38% | 46% | 6% |
| i | -4% | -13% | -8% | 10% | 10% | 0% | -13% |
| j | | | | | | | |
| k | -4% | -4% | -17% | 13% | 15% | -20% | -34% |
| l | -18% | -17% | -33% | 1% | 2% | -11% | -29% |
| m | -1% | -14% | -5% | 14% | 11% | -6% | -17% |
| n | -1% | -14% | 0% | 8% | 8% | -1% | -8% |
| o | - | - | - | - | - | - | - |
| p | - | - | - | - | - | - | - |
| q | - | - | - | - | - | - | - |
| r | -3% | -9% | -21% | 23% | 21% | -26% | -42% |
| s | -2% | -18% | -3% | 11% | 13% | 62% | 55% |
| t | -4% | -5% | -22% | 20% | 15% | -12% | -27% |

**FIGURES:**

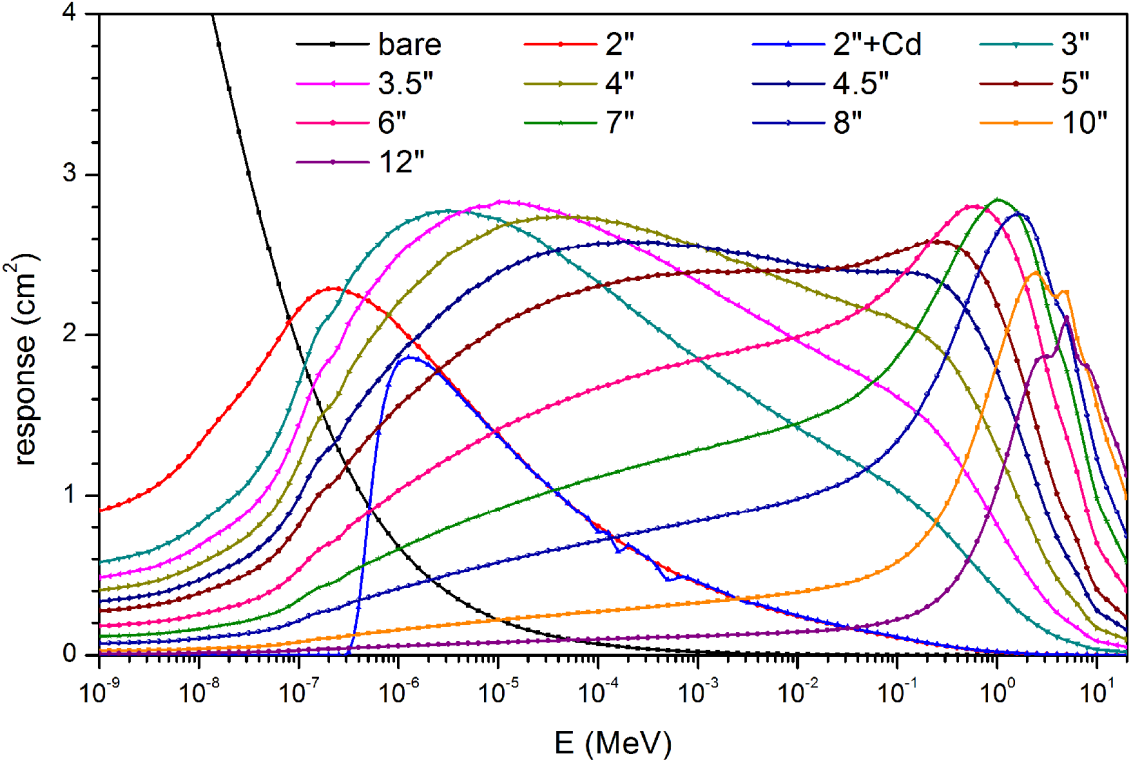

**Figure 1:** Calculated response matrix of the idealized ³He detector.

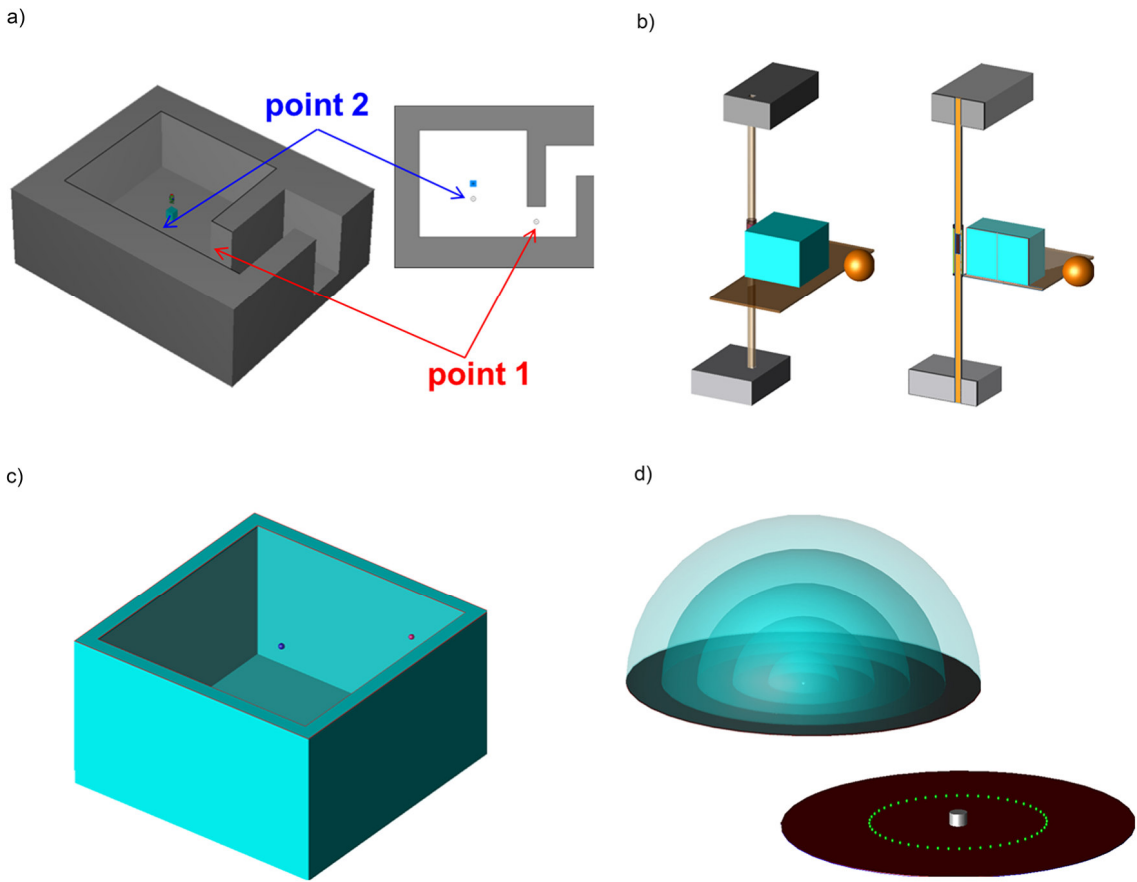

**Figure 2:** Irradiation scenarios: a) medical LINAC (2 measurement points); b) workplace; c) calibration facility; d) skyshine.

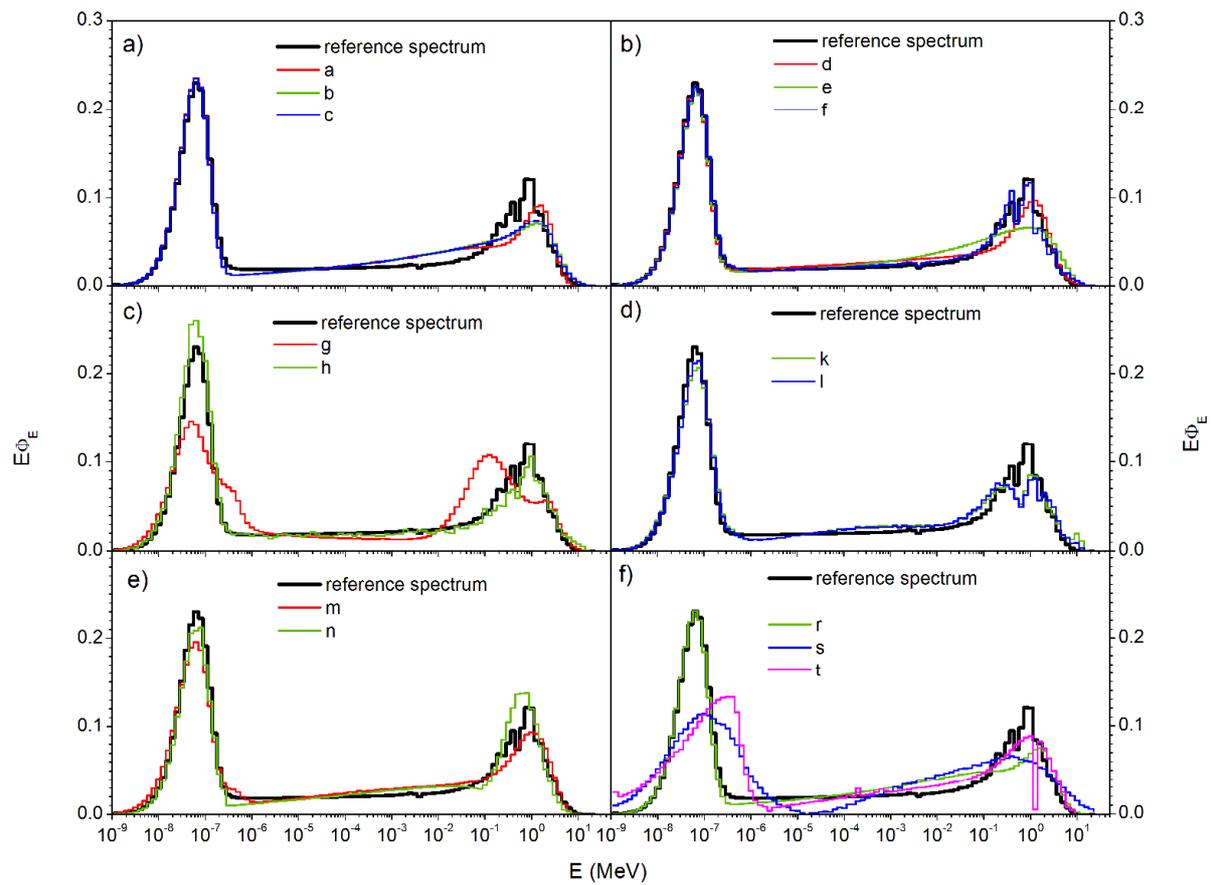

**Figure 3:** Participants unfolded spectra (in colour) compared with the reference spectra for the LINAC scenario, point 1 (at the entrance of the maze).

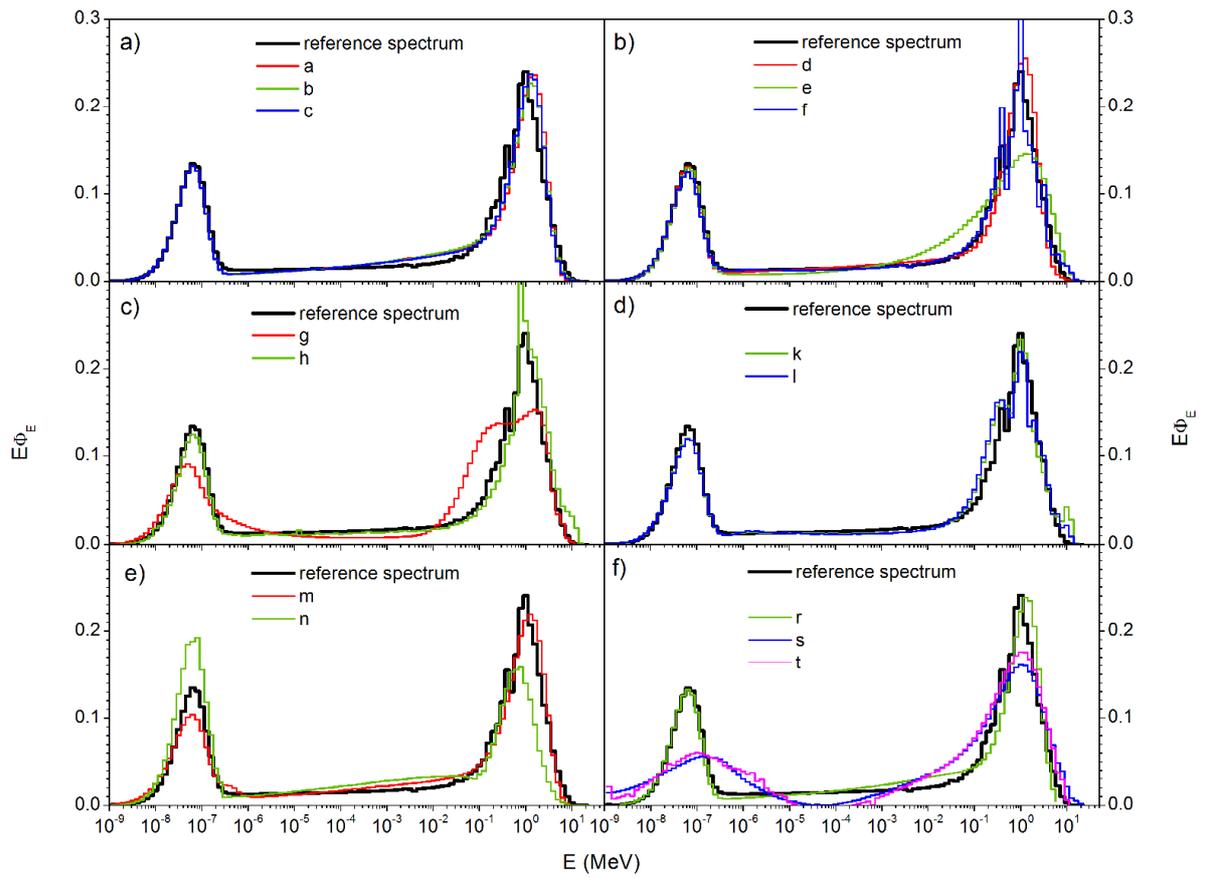

**Figure 4:** Participants unfolded spectra (in colour) compared with the reference spectra for the LINAC scenario, point 2 (1 m from the isocentre).

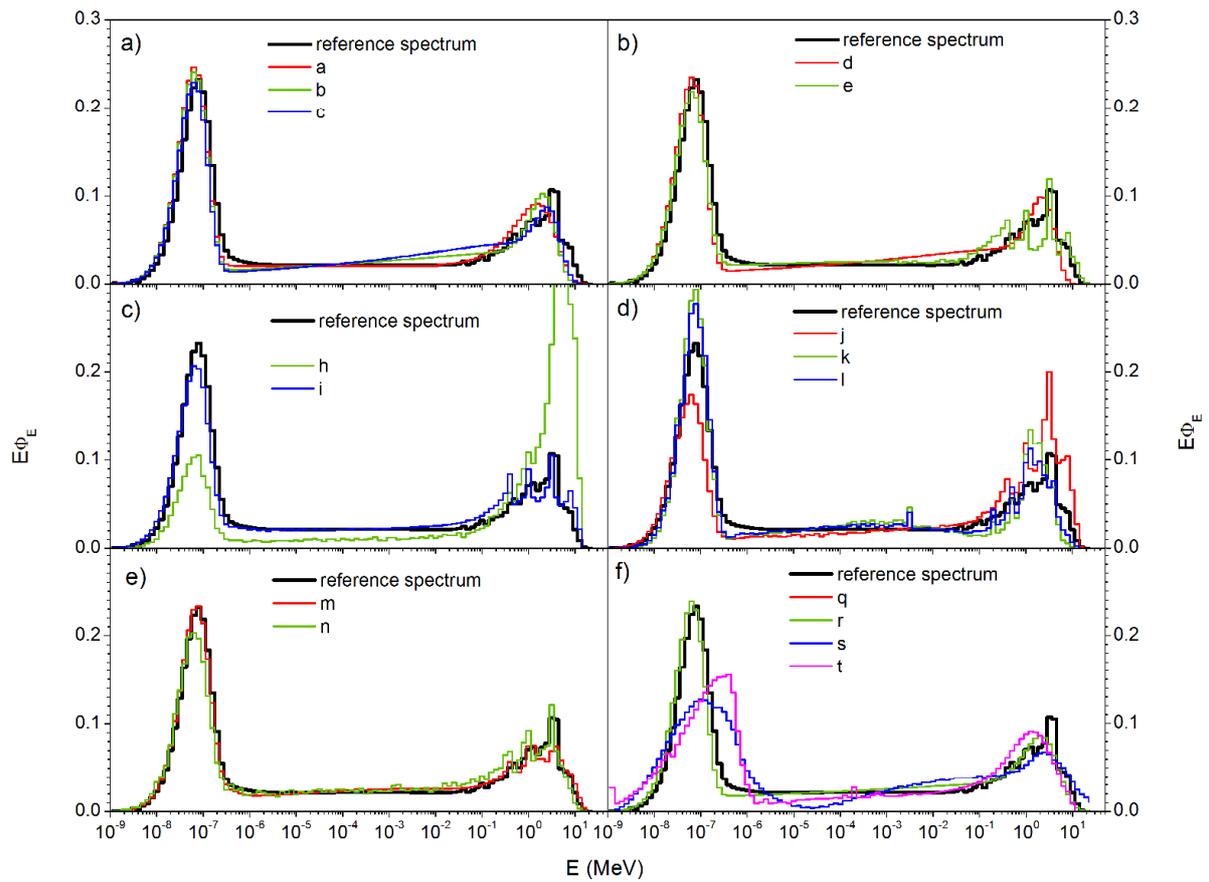

**Figure 5:** Participants unfolded spectra (in colour) compared with the reference spectra for the workplace scenario.

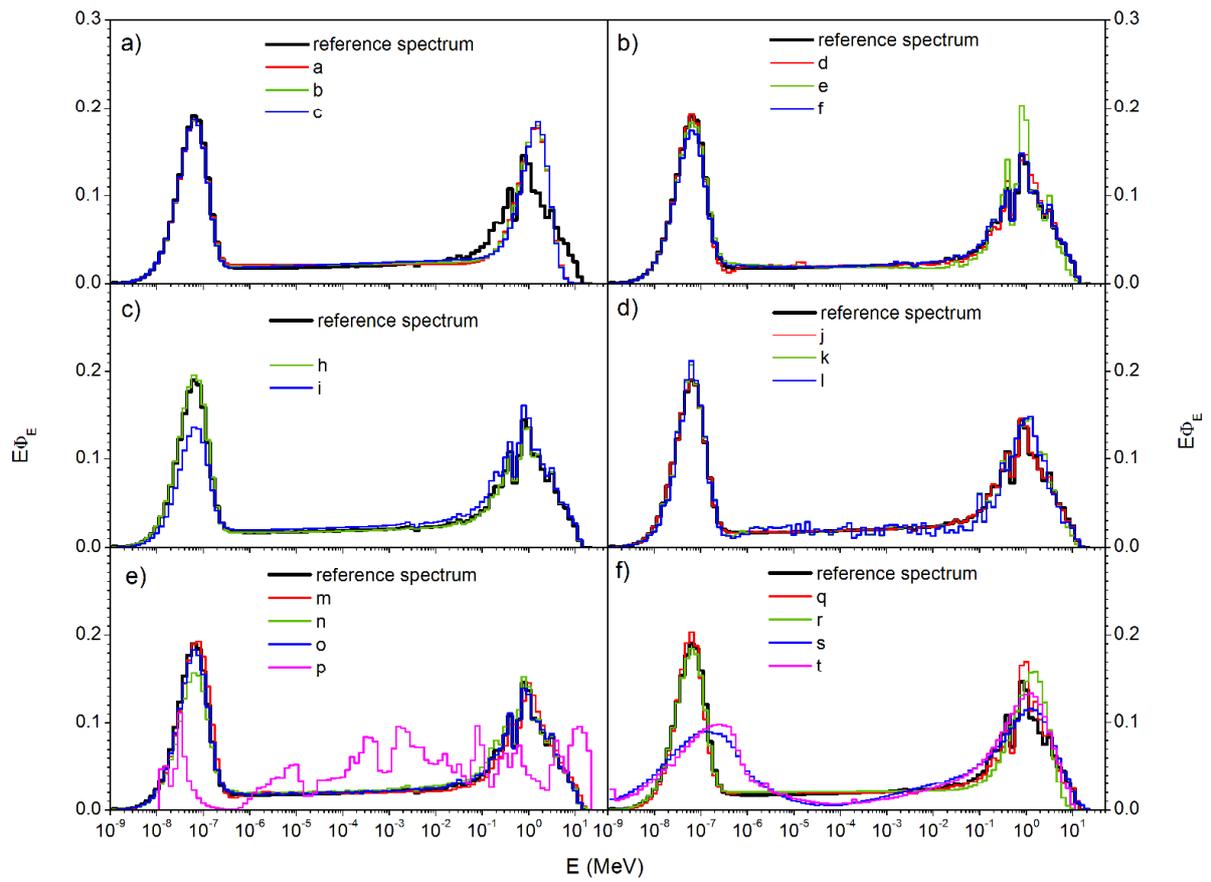

**Figure 6:** Participants unfolded spectra (in colour) compared with the reference spectra for the calibration facility.

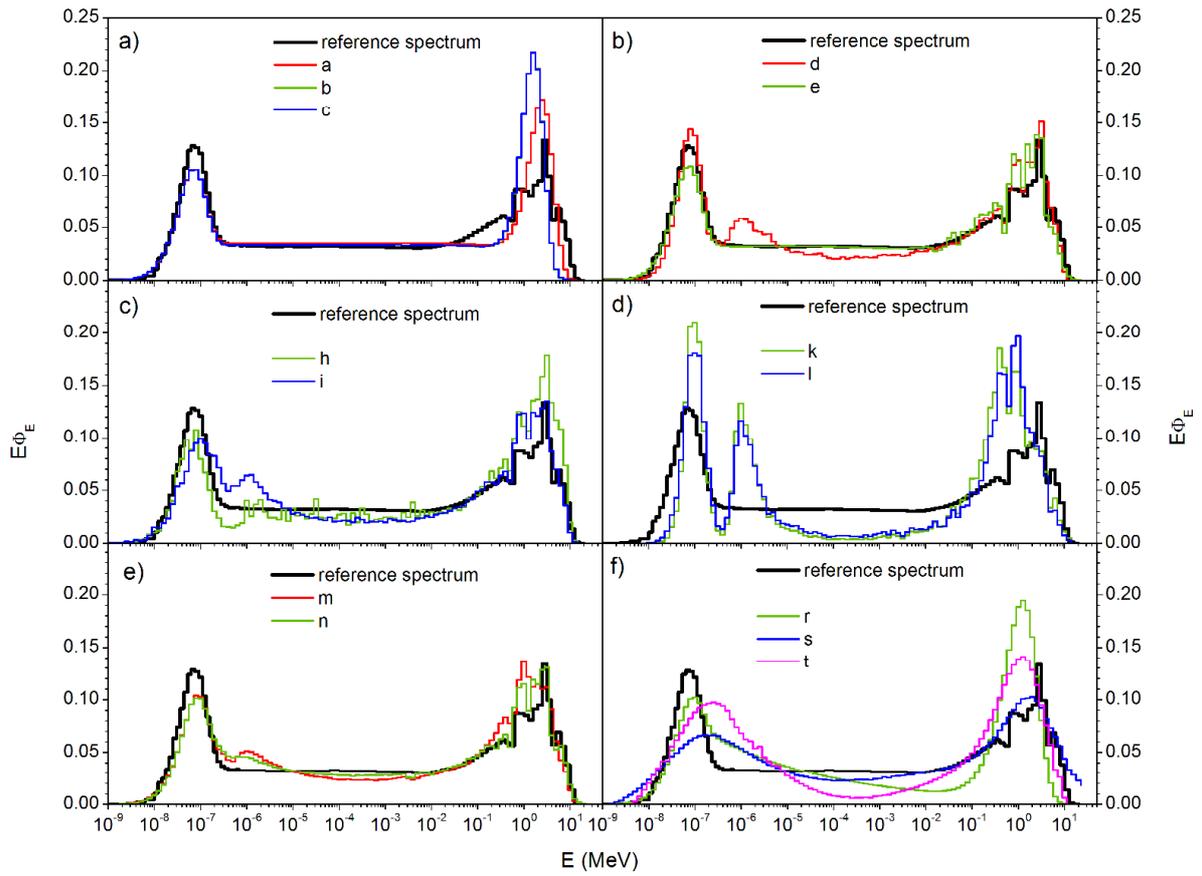

**Figure 7:** Participants unfolded spectra (in colour) compared with the reference spectra for the skyshine scenario.